# Manipulate Temperature Dependence of Thermal Conductivity of Graphene Phononic Crystal

Shiqian Hu[1], Meng An[2], Nuo Yang[2,3*], and Baowen Li[4,*]

[1]Center for Phononics and Thermal Energy Science, School of Physics Science and Engineering, Tongji University, Shanghai 200092, People's Republic of China

[2]Nano Interface Center for Energy (NICE), School of Energy and Power Engineering, Huazhong University of Science and Technology (HUST), Wuhan 430074, People's Republic of China

[3]State Key Laboratory of Coal Combustion, Huazhong University of Science and Technology (HUST), Wuhan 430074, People's Republic of China

[4]Department of Mechanical Engineering, University of Colorado, Boulder, CO 80309

*Corresponding authors:
Email: nuo@hust.edu.cn (N.Y.)
Email: Baowen.Li@Colorado.Edu (B.L.)

**Abstract**

By using non-equilibrium molecular dynamics simulations (NEMD), the modulation on temperature dependence of thermal conductivity of graphene phononic crystals (GPnCs) are investigated. It is found that the temperature dependence of thermal conductivity of GPnCs follows $\sim T^{-\alpha}$ behavior. The power exponents ($\alpha$) can be efficiently tuned by changing the characteristic size of GPnCs. The phonon participation ratio spectra and dispersion relation reveal that the long-range phonon modes are more affected in GPnCs with larger size of holes ($L_0$). Our results suggest that constructing GPnCs is an effective method to manipulate the temperature dependence of thermal conductivity of graphene, which would be beneficial for developing GPnCs-based thermal management and signal processing devices.

**Introduction**

Graphene is a two-dimensional hexagonal structure consisting of a single atomic layer of carbon.[1] Because of its excellent thermal, electrical, mechanical, and optical properties, graphene has attracted immense interest and been widely studied. Specifically, it has a superior thermal conductivity[2-6] which raises the exciting prospect of using graphene as a promising candidate for phononic (thermal) devices.[7-12]

Due to the heat transfer in graphene mainly contributed by phonons,[5] the thermal conductivity of graphene can be manipulated by the management of phonons. Traditionally, the thermal conductivity of graphene is manipulated through impurities,[13] superlattice structure,[14] or defects in the lattice crystal.[15-17]

Recently, there is a progress in managing phonons by nanostructured phononic crystals (PnCs)[18-23] which control heat by making use of phononic properties. It heralds the next technological revolution in phononics, such as thermal rectifiers,[15, 24-29] optomechanical crystals,[30, 31] thermal cloaking,[32-36] thermoelectrics,[37-41] and thermocrystals.[18, 21, 22] When the characteristic size of nanostructured PnCs is closed to the wavelength of phonons, PnCs may manipulate the phonon band structures which lead to the phonon confinement.[42-45] Moreover, the coherent interference is another underlying mechanism in manipulating phonons by PnCs.[22, 46, 47] These new findings are different from traditional methods in managing phonons and attracted a growing attention.

Efforts are spared to investigate the reducing of thermal conductivity of bulk materials by using PnCs structure. However, the temperature dependence of thermal conductivity of PnCs haven't been investigated systematically and are still confusing, which is important for both the fundamental researches and designing GPnCs-based thermal applications. In this letter, using NEMD simulations, the modulation on temperature dependence of thermal conductivity of GPnCs are investigated. Both the size of holes ($L_0$) and length (L) effect on thermal conductivity are studied.

**MD Simulation Methods**

As shown in Fig. 1(a), the GPnC is composed by periodically removing hexagonal holes whose edge is zigzag. The size of holes is characterized by $L_0$. The neck width ($L_1$) is fixed as 0.71 nm in all structures studied here. In calculations, the lattice constant, a, and thickness of graphene are chosen as 0.1418 nm and 0.334 nm, respectively. The fixed (periodic) boundary condition is adopted along longitudinal (transversal) direction.

In NEMD simulations, in order to establish a temperature gradient, the atoms at two ends are coupled with the Langevin heat bath,[48] whose temperature are $T_L$ and $T_R$ respectively. For convenience expression, we set the temperatures as $T_L=T_0(1+\triangle)$ and $T_R=T_0(1-\triangle)$, where $T_0$ is the average temperature, $\triangle$ is the normalized temperature difference between the two ends. In studying the dependence of thermal conductivity on temperature, $T_0$ ranges from 300 K to 1000 K, and $\triangle$ is fixed as 0.01.

The bonding interaction between carbon atoms is described by the energy potential a Morse bond and a harmonic cosine angle, which include both two-body and three-body potential terms.[26, 49, 50] We use the velocity Verlet algorithm to integrate the differential equations of motions, where the time step, $\Delta t$, is set as 0.5 fs. NEMD simulations are performed as 30 ns in the calculations of dependence of thermal conductivity on temperature after the system reaches a steady state.

The thermal conductivity ($\kappa$) are calculated based on the Fourier's Law as,

$$\kappa = -\frac{J}{A\nabla T} \quad (1),$$

where J is the heat current, A is the cross section and $\nabla T$ is the temperature gradient in the system. The results presented here are averaged from 12 independent simulations from different initial conditions. The error bar is the standard deviations of the results of 12 simulations.

**MD Results and Discussions**

Firstly, the temperature dependence of thermal conductivity is investigated in several GPnC structures whose L (W) is fixed as 25 nm (7 nm), and $L_0$ are increased from 0.71 nm to 2.41 nm. Fig. 1(b) shows a typical temperature profile of GPnC. The jumps at the two ends come from the thermal interfacial resistance between heat baths and system. Excluding the effects at two ends, the data is fitted linearly to obtain the temperature gradient.

The results of temperature dependence of thermal conductivity are shown in Fig. 2(a). For the convenience of comparison, Fig. 2(a) show the thermal conductivity of both graphene ($\kappa_G$) and GPnCs ($\kappa_{GPnCs}$) with the same length, L, as 25 nm. It is found that both $\kappa_G$ and $\kappa_{GPnCs}$ decrease with the temperature increasing as $\sim T^{-\alpha}$ due to the Umklapp phonon-phonon scattering. Graphene has a $T^{-1}$ behavior followed Slack relation,[51] which is consistent with previous work.[45] However, GPnCs have different power exponent, $\alpha$, corresponding to its size of holes, $L_0$. The results show that the values of $\alpha$ are 0.69, 0.51, 0.41, 0.34 and 0.28 corresponding to $L_0$ as 0.71, 1.13, 1.55, 1.98 and 2.41 nm, respectively. Namely, the value of $\alpha$ decreases with $L_0$ increasing (shown in Fig. 2(b)). It is known that the Slack relation is valid only when the long-range acoustic phonons dominate the thermal transport, meanwhile, both short-range acoustic and optical phonons are not important.[51, 52] Sometimes, this precondition is not satisfied for structures with low thermal conductivity, such as zeolites[53] and GPnCs in our work.

As shown in Fig. 2(a), compared with the thermal conductivity of graphene, $\kappa_{GPnCs}$ are significantly decrease and have a less dependence on temperature. It has been demonstrated that the thermal conductivity contributed by short-range acoustic and optical phonons is temperature independent in previous works.[53, 54] In graphene, the long-range acoustic phonons dominate the thermal transport.[55] The significantly decrease of thermal conductivity of GPnCs and a less sensible dependence on temperature should be caused by the affected long-range acoustic phonons.

Moreover, as shown in Fig. 2(a), $\kappa_{GPnCs}$ decrease with $L_0$ increasing in a large temperature range (300 K to 1000 K). For instance, when $L_0$ equals 2.41 nm, $\kappa_{GPnC}$ is reduced to 4% of $\kappa_G$ at room temperature. The $\kappa_{GPnCs}$ with different $L_0$ at room temperature are shown in Fig. 2(c). The lower group velocities due to the flatted phonon dispersion curves (details in Fig. 4(b) and (c)) and phonon modes more localized in GPnC with larger $L_0$ (details in Fig .4(a)) should be responsible for its lower thermal conductivity.

Besides the size of holes ($L_0$), the length of GPnCs (L) is another characteristic length affecting the temperature dependence of $\kappa_{GPnCs}$. We fix $L_0$ as 0.71 nm and meanwhile change L from 7 nm to 31 nm. The thermal conductivity results are shown in Fig. 3(a). It is found that $\kappa_{GPnCs}$ also follow ~ $T^{-\alpha}$ behavior. The values of α are 0.39, 0.54, 0.63, 0.69 and 0.72 corresponding L as 7, 14, 19, 25 and 31 nm, respectively. Namely, the power exponents (α) decreases with L decreasing (shown in Fig. 3(b)). Due to finite size effects, the phonons, whose wavelength is larger than L, cannot exist and their contribution to thermal conductivity is reduced. With the length (L) decreasing, the short-range acoustic or optical phonons contributed to $\kappa_{GPnCs}$ increases, which leads to the reduction of the value of α [53, 54]. This result indicates that the value of α could be effectively tuned by changing the length of GPnCs.

Besides, as shown in Fig. 3(c), $\kappa_{GPnCs}$ diverge with L increasing as log(L) behavior in a large temperature range (300 K to 1000 K). It exhibits the same behavior as $\kappa_G$ which has been found in the experimental study.[5] This result shows a consequence of the two-dimensional nature of phonons in GPnCs, and provides strong evidence that Fourier's law of heat conduction is not valid in two-dimensional nanostructures just as in one-dimensional nanostructures.[56]

In order to understand the underlying physical mechanism of the reduction of power exponent, α, the participation ratio spectra and dispersion relation of graphene and GPnCs are calculated (shown in Fig. 4). The participation ratio (P) of phonon mode k is defined by the normalized eigenvector $\varepsilon_{i\alpha,k}$

$$p_k = \frac{1}{N\sum_{i=1}^{N}(\sum_{\alpha=1}^{3}\varepsilon^*_{i\alpha,k}\varepsilon_{i\alpha,k})^2} \qquad (2),$$

where N is the total number of atoms, $\varepsilon_{i\alpha,k}$ is calculated by general utility lattice program (GULP).[57] When there are less atoms participating in the motion, the phonon mode has a smaller value of P.

As shown in Fig. 4(a), the participation ratios of GPnCs are significantly reduced compared with that of graphene, thus phonon modes of GPnCs are more localized (a smaller relaxation time) than graphene. Additionally, there exists a further reduction of the participation ratios in GPnCs when $L_0$ changes from 0.71 nm to 2.41 nm, which means that larger $L_0$ could enhance phonon localizations in GPnCs. Namely, with $L_0$ increasing, the phonon relaxation time of GPnCs decrease. Fig. 4(b) and (c) show the lower-frequency part of phonon dispersion of graphene and GPnC. The phonon dispersions of graphene and GPnC are calculated by lattice dynamics implemented in GULP.[57] Obviously, the phonon dispersions are folded and flattened in GPnC, thus the group velocity are significantly decrease, especially for the acoustic phonons. In GPnCs, combining the phonon relaxation time and group velocity effect, with the $L_0$ increase, the long-range acoustic phonons will be suppressed, thus the relative contribution from the short-range acoustic and optical phonons to $\kappa_{GPnC}$ increases compared with that in the graphene case, which violates the valid condition of the Slack relation and leads to a less sensible dependence on temperature.

**Summary**

In summary, we have studied the modulation on temperature dependence of thermal conductivity of GPnCs by using NEMD simulations. The results show that, compared to graphene, the thermal conductivity of GPnCs have a less dependence on temperature. There is a relationship as $\kappa_{GPnCs} \sim T^{-\alpha}$ where the value of $\alpha$ can be tuned by the characteristic length of GPnCs, such as $L_0$ and L. Our results demonstrate that constructing GPnCs structure is an effective way to tune the temperature dependence of $\kappa_{GPnCs}$. The ability to tune and control the temperature dependence of thermal conductivity of GPnCs provides potential ways to thermal design of GPnCs-based signal devices at the nanoscale.

**Acknowledgements**

This project was supported in part by the grants from the National Natural Science Foundation of China: 11334007 (BL), 51576076 (NY) and 11204216 (NY).We are grateful to Nianbei Li, Qichen Song, Dengke Ma, Zelin Jin, Weiwei Zhu, Biao Wang, Tingyu Lu for useful discussions. The authors thank the National Supercomputing Center in Tianjin (NSCC-TJ) for providing assistance in computations.

**References**

[1] K. Novoselov, A. K. Geim, S. Morozov, D. Jiang, M. K. I. Grigorieva, S. Dubonos, and A. Firsov, Nature **438** 197 (2005).

[2] A. A. Balandin, S. Ghosh, W. Bao, I. Calizo, D. Teweldebrhan, F. Miao, and C. N. Lau, Nano Lett **8** 902 (2008).

[3] S. Ghosh, I. Calizo, D. Teweldebrhan, E. P. Pokatilov, D. L. Nika, A. A. Balandin, W. Bao, F. Miao, and C. N. Lau, Appl Phys Lett **92** 151911 (2008).

[4] J.-U. Lee, D. Yoon, H. Kim, S. W. Lee, and H. Cheong, Phys Rev B **83** 081419 (2011).

[5] X. Xu, L. F. Pereira, Y. Wang, J. Wu, K. Zhang, et al., Nat Commun **5** 3689 (2014).

[6] B. D. Kong, S. Paul, M. B. Nardelli, and K. W. Kim, Phys Rev B **80** 033406 (2009).

[7] B. W. Li, L. Wang, and G. Casati, Appl Phys Lett **88** 143501 (2006).

[8] L. Wang and B. Li, Phys Rev Lett **99** 177208 (2007).

[9] B. Q. Ai, W. R. Zhong, and B. B. Hu, J Phys Chem C **116** 13810 (2012).

[10] B. Q. Ai, M. An, and W. R. Zhong, J Chem Phys **138** 034708 (2013).

[11] P. Ben-Abdallah and S. A. Biehs, Phys Rev Lett **112** 044301 (2014).

[12] H. Han, B. Li, S. Volz, and Y. A. Kosevich, Phys Rev Lett **114** 145501 (2015).

[13] S. Chen, Q. Wu, C. Mishra, J. Kang, H. Zhang, K. Cho, W. Cai, A. A. Balandin, and R. S. Ruoff, Nat Mater **11** 203 (2012).

[14] T. Ouyang, Y. P. Chen, K. K. Yang, and J. X. Zhong, Epl **88** 28002 (2009).

[15] J. Hu, X. Ruan, and Y. P. Chen, Nano Lett **9** 2730 (2009).

[16] J. Haskins, A. Kınacı, C. Sevik, H. Sevinçli, G. Cuniberti, and T. Çağın, ACS nano **5** 3779 (2011).

[17] Z. G. Fthenakis, Z. Zhu, and D. Tománek, Phys Rev B **89** 125421 (2014).

[18] J. K. Yu, S. Mitrovic, D. Tham, J. Varghese, and J. R. Heath, Nat Nanotechnol **5** 718 (2010).

[19] J. Tang, H. T. Wang, D. H. Lee, M. Fardy, Z. Huo, T. P. Russell, and P. Yang, Nano Lett **10** 4279 (2010).

[20] P. E. Hopkins, C. M. Reinke, M. F. Su, R. H. Olsson, E. A. Shaner, Z. C. Leseman, J. R. Serrano, L. M. Phinney, and I. El-Kady, Nano Lett **11** 107 (2011).

[21] M. Maldovan, Nature **503** 209 (2013).

[22] M. Maldovan, Phys Rev Lett **110** 025902 (2013).

[23] J. Lee, J. Lim, and P. Yang, Nano Lett **15** 3273 (2015).

[24] N. B. Li, J. Ren, L. Wang, G. Zhang, P. Hanggi, and B. W. Li, Rev Mod Phys **84** 1045 (2012).

[25] N. Yang, N. Li, L. Wang, and B. Li, Phys Rev B **76** 020301 (2007).

[26] N. Yang, G. Zhang, and B. W. Li, Appl Phys Lett **93** 243111 (2008).

[27] N. Yang, G. Zhang, and B. W. Li, Appl Phys Lett **95** 033107 (2009).

[28] B. Li, L. Wang, and G. Casati, Phys Rev Lett **93** 184301 (2004).

[29] C. Chang, D. Okawa, A. Majumdar, and A. Zettl, Science **314** 1121 (2006).

[30] M. Eichenfield, J. Chan, R. M. Camacho, K. J. Vahala, and O. Painter, Nature **462** 78 (2009).

[31] A. H. Safavi-Naeini, J. T. Hill, S. Meenehan, J. Chan, S. Groblacher, and O. Painter, Phys Rev Lett **112** 153603 (2014).

[32] T. Han, X. Bai, J. T. Thong, B. Li, and C. W. Qiu, Adv Mater **26** 1731 (2014).
[33] S. Narayana, S. Savo, and Y. Sato, Appl Phys Lett **102** 201904 (2013).
[34] S. Guenneau, C. Amra, and D. Veynante, Opt Express **20** 8207 (2012).
[35] T. Han, X. Bai, D. Gao, J. T. Thong, B. Li, and C. W. Qiu, Phys Rev Lett **112** 054302 (2014).
[36] T. Han, T. Yuan, B. Li, and C. W. Qiu, Sci Rep **3** 1593 (2013).
[37] G. Chen, M. S. Dresselhaus, G. Dresselhaus, J. P. Fleurial, and T. Caillat, Int Mater Rev **48** 45 (2003).
[38] M. S. Dresselhaus, G. Chen, M. Y. Tang, R. G. Yang, H. Lee, D. Z. Wang, Z. F. Ren, J. P. Fleurial, and P. Gogna, Adv Mater **19** 1043 (2007).
[39] C. J. Vineis, A. Shakouri, A. Majumdar, and M. G. Kanatzidis, Adv Mater **22** 3970 (2010).
[40] Z. Jin, Q. Liao, H. Fang, Z. Liu, W. Liu, Z. Ding, T. Luo, and N. Yang, arXiv preprint arXiv:1504.03852 (2015).
[41] L. Yang, N. Yang, and B. Li, arXiv preprint arXiv:1410.8193 (2014).
[42] A. Balandin and K. L. Wang, Phys Rev B **58** 1544 (1998).
[43] L. Yang, N. Yang, and B. Li, Sci Rep **3** 1143 (2013).
[44] L. Yang, N. Yang, and B. Li, Nano Lett **14** 1734 (2014).
[45] L. Yang, J. Chen, N. Yang, and B. Li, arXiv preprint arXiv:1407.5885 (2014).
[46] N. Zen, T. A. Puurtinen, T. J. Isotalo, S. Chaudhuri, and I. J. Maasilta, Nat Commun **5** 3435 (2014).
[47] S. Alaie, D. F. Goettler, M. Su, Z. C. Leseman, C. M. Reinke, and I. El-Kady, Nat Commun **6** 7228 (2015).
[48] N. Yang, G. Zhang, and B. Li, Nano Lett **8** 276 (2008).
[49] Y. Quo, N. Karasawa, and W. A. Goddard, Nature **351** 464 (1991).
[50] R. E. Tuzun, D. W. Noid, B. G. Sumpter, and R. C. Merkle, Nanotechnology **7** 241 (1996).
[51] G. A. Slack, Solid state physics **34** 1 (1979).
[52] M.Kaviany, Cambridge University Press, New York p.212 (2008).
[53] A. McGaughey and M. Kaviany, Int J Heat Mass Tran **47** 1799 (2004).
[54] A. J. H. McGaughey and M. Kaviany, Int J Heat Mass Tran **47** 1783 (2004).
[55] T. L. Feng, X. L. Ruan, Z. Q. Ye, and B. Y. Cao, Phys Rev B **91** 224301 (2015).
[56] N. Yang, G. Zhang, and B. Li, Nano Today **5** 85 (2010).
[57] J. D. Gale, J Chem Soc Faraday T **93** 629 (1997).

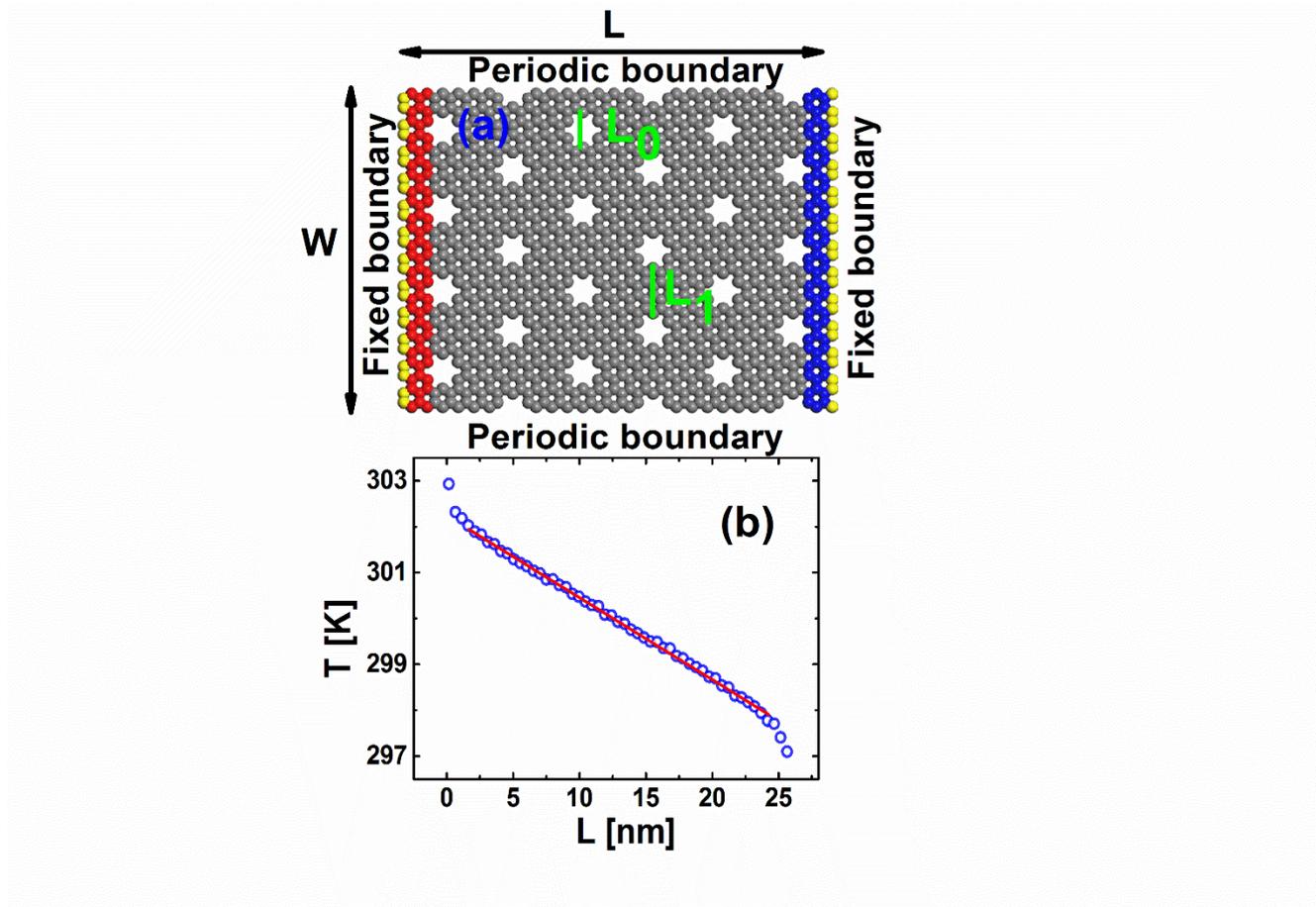

Figure 1. (Color on-line) (a) Schematic picture of graphene phononic crystal (GPnC). The length (width) of the GPnC is set as L (W). $L_0$ (size of holes) and $L_1$ (neck width) are used to characterize the GPnC structure. $L_1$ is fixed as 0.71 nm. (b) The temperature profile of the GPnC. The parameters are L=25 nm, W=7 nm, $L_0$=0.71 nm, $T_0$=300 K and $\triangle$ = 0.01. The temperature gradient is obtained by fitting data excluding the temperature jump at two ends.

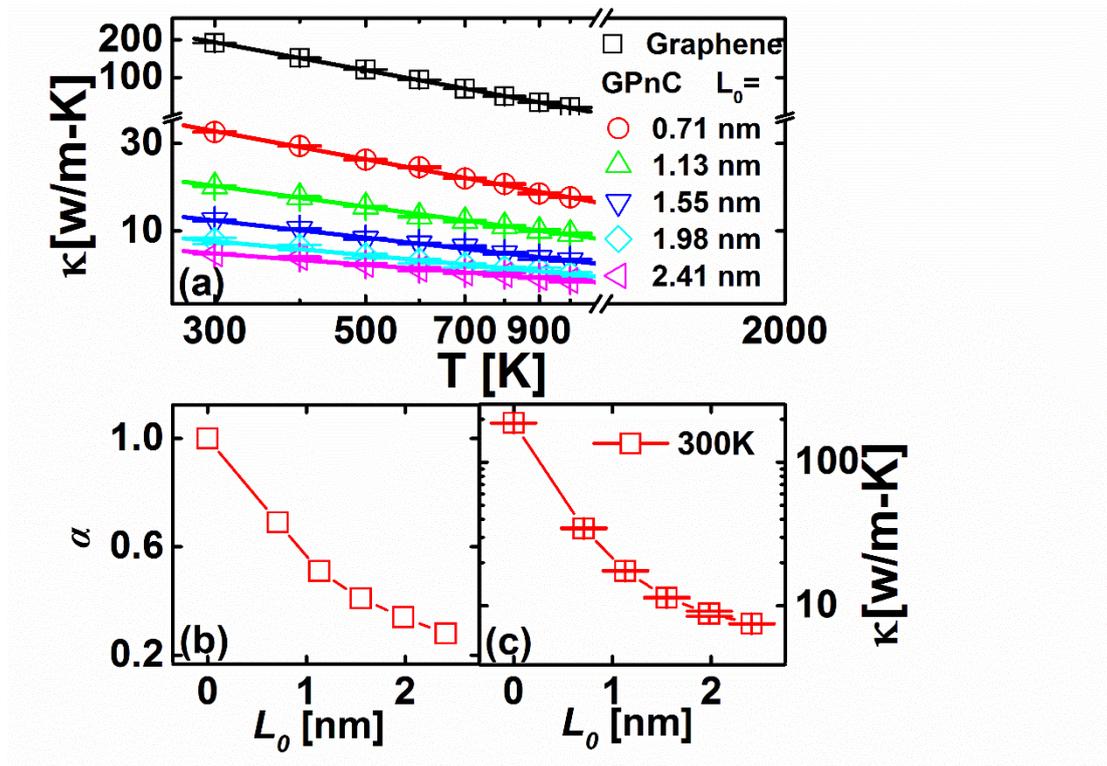

Figure 2. (Color on-line) (a) Thermal conductivity of graphene and GPnCs with different $L_0$ versus the temperature. The temperature increases from 300 K to 1000 K. The parameters are L=25 nm, W=7 nm and $\triangle$ = 0.01. For GPnCs, $L_1$=0.71 nm and $L_0$ increases from 0.71 nm to 2.41 nm. The figure is plotted in log-log scale. The symbols are numerical data and the lines are fitted lines. The fitted values α are 0.69, 0.51, 0.41, 0.34, 0.28 corresponding $L_0$ as 0.71, 1.13, 1.55, 1.98, 2.41 nm, respectively. Especially, for graphene, α equals 1. (b) The power exponents (α) versus $L_0$. The values of α decrease with $L_0$ increasing. (c) Thermal conductivity of GPnCs versus $L_0$ at 300 K. The error bar is standard deviation of 12 MD simulations with different initial conditions.

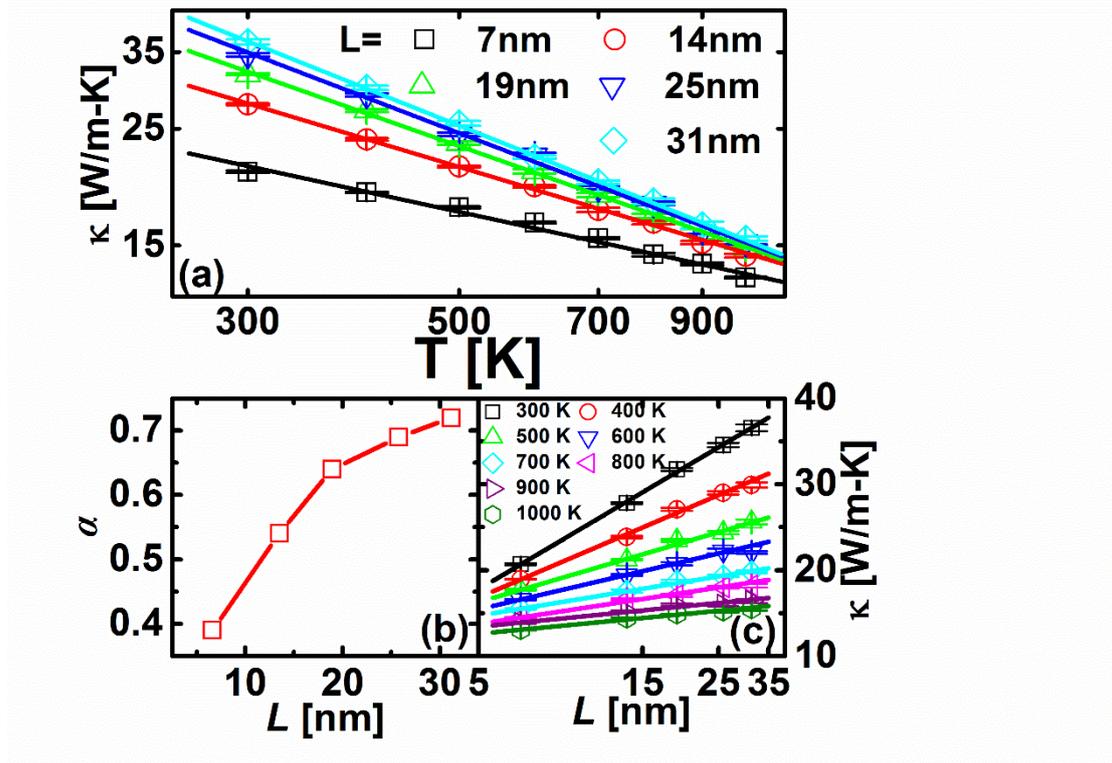

Figure 3. (Color on-line) (a) Thermal conductivity of GPnCs with different lengths L versus the temperature. The temperature increases from 300 K to 1000 K. The parameters W =7 nm, $L_1$=0.71 nm, $L_0$= 0.71 nm and △= 0.01. The length L increases from 7 nm to 31 nm. The figure is plotted in log-log scale. The symbols are numerical data and the lines are fitted lines. The fitted values α are 0.39, 0.54, 0.63, 0.69 and 0.72 corresponding L as 7, 14, 19, 25 and 31 nm, respectively. (b) The power exponents (α) versus L. The values of α decrease with L decreasing. (c) Thermal conductivity of GPnCs versus L at different temperatures. The symbols are numerical data and the dash lines are fitted lines. The error bar is standard deviation of 12 MD simulations with different initial conditions.

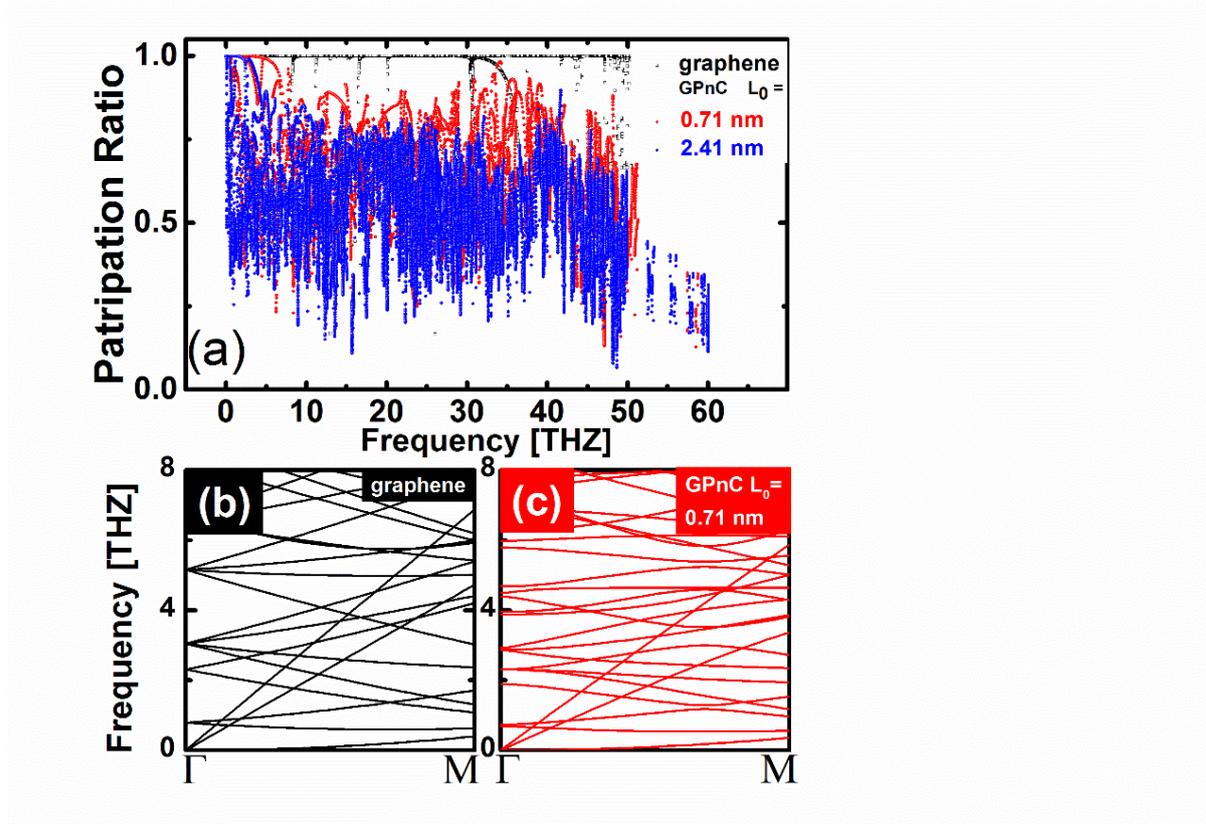

Figure 4. (Color on-line) (a) Participation ratio spectra of graphene and GPnCs with different $L_0$ (0.71 nm or 2.41 nm). Black points correspond to participation ratios of graphene. Red points and blue points correspond to participation ratio of GPnCs with $L_0$ 0.71 nm and 2.41 nm, respectively. GPnCs have smaller participation ratio than graphene. Additionally, GPnC with smaller $L_0$ has larger participation ratio. (b) Low frequency part of the phonon dispersion of graphene. (c) Low frequency part of phonon dispersion of GPnC. The square graphene cell in (b) and the GPnC in (c) are the same as the graphene cell in (a), and the size of holes $L_0$ in GPnC is 0.71 nm. Phonon dispersions in GPnC are flattened compared with that in graphene.